\documentclass[conference]{IEEEtran}
\IEEEoverridecommandlockouts
\usepackage{cite}
\usepackage{amsmath,amssymb,amsfonts}
\usepackage{algorithmic}
\usepackage{graphicx}
\usepackage{textcomp}
\usepackage{xcolor}
\usepackage[english]{babel}
\usepackage{tikz}
\usepackage{xcolor}

\usepackage{color}

\usepackage{booktabs}
\usepackage{multirow}
\usepackage{graphicx}
\usepackage{colortbl}
\def\BibTeX{{\rm B\kern-.05em{\sc i\kern-.025em b}\kern-.08em
    T\kern-.1667em\lower.7ex\hbox{E}\kern-.125emX}}
\begin{document}

\title{Mindful-RAG: A Study of Points of Failure in Retrieval Augmented Generation}


\author{
\IEEEauthorblockN{Garima Agrawal,
Tharindu Kumarage,
Zeyad Alghamdi, and 
Huan Liu}
\IEEEauthorblockA{School of Computing and Augmented Intelligence, Arizona State University, Tempe, USA \\
Email: \{garima.agrawal, kskumara, zalgham1, huanliu\}@asu.edu}
}


\maketitle

\begin{abstract}
Large Language Models (LLMs) excel at generating coherent text but often struggle with knowledge-intensive queries, particularly in domain-specific and factual question-answering tasks. Retrieval-augmented generation (RAG) systems have emerged as a promising solution by integrating external knowledge sources, such as structured knowledge graphs (KGs). While KG-based RAG approaches have demonstrated value, current state-of-the-art solutions frequently fall short, failing to deliver accurate and reliable answers even when the necessary factual knowledge is available. In this paper, we present a critical analysis of failure points in existing KG-based RAG methods, identifying eight key areas of concern, including misinterpretation of question context, incorrect relation mapping, and ineffective ambiguity resolution. We argue that these failures primarily stem from design limitations in current KG-RAG systems, such as inadequate attention to discerning user intent and insufficient alignment of retrieved knowledge with the contextual demands of the query. Based on this analysis, we propose a new approach for KG-RAG systems, termed Mindful-RAG, which re-engineers the retrieval process to be more intent-driven and contextually aware. By enhancing reasoning capabilities, improving constraint identification, and addressing the structural limitations of knowledge graphs, we aim to improve the reliability and effectiveness of KG-RAG systems. To validate this approach, we developed a proof-of-concept by integrating the principles of Mindful-RAG into an existing KG-RAG system. The Mindful-RAG approach seeks to deliver more robust, accurate, and contextually aligned AI-driven knowledge retrieval systems, with potential applications in critical domains such as healthcare, legal, research, and scientific discovery, where precision and reliability are paramount.
\end{abstract}


\section{Introduction}
Large Language Models (LLMs) have revolutionized natural language processing, excelling in various tasks. However, they frequently generate hallucinated responses when dealing with domain-specific or knowledge-intensive queries \cite{ji2023survey}. This limitation has led to the development of Retrieval-augmented Generation (RAG) methods, which enable LLMs to access and incorporate external knowledge sources, such as structured knowledge graphs (KGs) \cite{li2024enhancing, ding2024survey}.

Despite the promise of RAG methods, particularly those integrating KGs, significant challenges persist. Even with access to relevant information, these systems often fail to provide accurate answers, especially as query complexity increases \cite{gao2023retrieval, agrawal2023can, jeong2024adaptive}. To better understand these limitations, We conducted a study to critically analyze the failure points in existing KG-based RAG methods. Our investigation identified eight critical failure points in these systems, which we categorized into two primary areas:
\begin{enumerate} 
    \item \textbf{Reasoning Failures}: LLMs struggle to accurately interpret user queries and leverage contextual information, resulting in a misalignment between retrieved knowledge and query intent.
    \item \textbf{Structural Limitations}: These failures primarily arise from insufficient attention to the structure of knowledge sources, such as knowledge graphs, and the use of inappropriate evaluation metrics. 
\end{enumerate}

Given these persistent issues, there is a pressing need to look beyond conventional approaches and critically reassess how KG-RAG systems are designed. To address these challenges, we propose \textbf{Mindful-RAG}, an approach that re-engineers the retrieval process to be more intent-driven and contextually aware. Mindful-RAG is not merely an alternative method; it represents a comprehensive approach aimed at the development of more effective KG-RAG systems. Unlike traditional methods that primarily rely on semantic similarity or structural cues, Mindful-RAG suggests to leverage the intrinsic parametric knowledge of LLMs to accurately discern the intent behind queries. This approach not only guides the retrieval process to ensure that the extracted context from the KG is relevant but also aligns it with the original intent of the query. Additionally, Mindful-RAG introduces advanced contextual alignment techniques for efficient knowledge graph navigation and incorporates a validation step to ensure the generated response meets the intended requirements. To validate this approach, we developed a proof-of-concept that integrates Mindful-RAG into an existing KG-RAG system through prompt engineering. Preliminary experiments on WebQSP and MetaQA datasets indicate promising results compared to existing state-of-the-art methods, particularly in reducing reasoning errors by enhancing the focus on understanding query objectives and improving contextual alignment.
In this paper, we make the following key contributions: \begin{enumerate}
    \item A comprehensive error analysis of KG-based RAG methods, identifying eight critical failure points and highlighting design limitations in state-of-the-art frameworks, particularly in addressing question intent and achieving contextual alignment in vanilla-RAG systems. 
    \item The introduction of a proof-of-concept that opens a novel research direction, redefining the RAG pipeline by leveraging LLMs' parametric memory for enhanced intent identification and contextual alignment. 
\end{enumerate}
By addressing these fundamental challenges, Mindful-RAG aims to create more robust, accurate, and contextually aligned AI-driven knowledge retrieval systems, particularly in precision-critical fields such as healthcare, legal, research, and scientific discovery.

\section{KG-based RAG Failure Analysis}
\begin{table*}[htbp]
\centering
\caption{KG-Based RAG Failure Analysis}
\label{tab:err_ana}

\definecolor{customgrey}{rgb}{0.7, 0.7, 0.7}

\resizebox{\textwidth}{!}{%
\begin{tabular}{>{\raggedright\arraybackslash}p{2.5cm}>{\raggedright\arraybackslash}p{3cm}>{\raggedright\arraybackslash}p{5cm}>{\raggedright\arraybackslash}p{7cm}}
\toprule
\rowcolor{gray!30}
\textbf{Error Category} & \textbf{Error Type} & \textbf{Description} & \textbf{Representative Failed Example(s)} \\
\midrule
\multirow{5}{*}[-2ex]{\textbf{Reasoning Failures}} & 
\cellcolor{customgrey}Misinterpretation of Question's Context & 
LLMs misinterpret the question or fail to understand specific requirements of the question. & 
\begin{itemize}
    \item Failed to relate \textbf{Justin Bieber's} birthplace to his country of birth, focusing on city-level information instead of the required higher geographical context.
    \item Incorrectly identified the location of \textbf{Fukushima Daiichi} nuclear plant, choosing the city \textbf{'Fukushima'} instead of the correct town \textbf{'Okuma'} and country \textbf{'Japan'}.
\end{itemize} \\
\cmidrule{2-4}
& \cellcolor{customgrey}Incorrect Relation Mapping & 
LLMs often choose relations that do not correctly address the question. & 
For a question about where \textbf{Andy Murray} started playing tennis, choosing \textit{people.person.place\_of\_birth} suggests a misunderstanding of the question's intent. \\
\cmidrule{2-4}
& \cellcolor{customgrey}Ambiguity in Question or Data & 
LLMs fail to identify key terms and their meanings or implications across various contexts from the provided KG triples. & 
\begin{itemize}
    \item Could not identify the \textbf{Serbian language} from the list of languages spoken in \textbf{Serbia}.
    \item Failed to recognize that a query was about the \textbf{"most" exported item}, not just any exported item.
\end{itemize} \\
\cmidrule{2-4}
& \cellcolor{customgrey}Specificity or Precision Errors & 
LLMs often misinterpret questions requiring aggregated responses as specific, singular answers. They also struggle with temporal context. & 
\begin{itemize}
    \item Picked \textbf{2000} as \textbf{George W. Bush's} election year without considering his two elections (\textbf{2000 and 2004}).
    \item Selected \textbf{'Sue Douglas'} as \textbf{Niall Ferguson's} spouse instead of finding the current spouse, \textbf{'Ayaan Hirsi Ali'}, ignoring multiple spouse possibilities.
\end{itemize} \\
\cmidrule{2-4}
& \cellcolor{customgrey}Constraint Identification Error & 
LLMs fail to correctly identify or apply constraints provided or implied in the question. & 
\begin{itemize}
    \item Could not effectively narrow the search for \textbf{Jackie Robinson's} first team.
    \item For "Who played Bilbo in Lord of the Rings?", LLMs identified \textbf{"Old Bilbo"} and specific films but failed to derive a single definitive answer.
\end{itemize} \\
\midrule
\multirow{3}{*}[-1ex]{\textbf{Structural Limitation}} & 
\cellcolor{customgrey}Encoding Issues & 
Compound value types (CVTs) in KGs represent complex data. If mismanaged or unrecognized by LLMs, they may be misinterpreted as final answers. & 
For 'Where is the Sony Ericsson Company?', the model correctly identifies relations but mistakenly selects CVT\_0 as the final answer due to CVT node linking. \\
\cmidrule{2-4}
& \cellcolor{customgrey}Inappropriate Evaluation & 
The exact match (EM) module only accepts fully correct answers and sometimes fails due to misinterpreting the required depth of information or mis-aligning with expected answer format. & 
For 'What year did the Orioles go to the World Series?', the model retrieved correct years (1983, 1970, 1966) but failed to match the expected format \textbf{[1983 World Series, 1970 World Series, 1966 World Series]}. \\
\cmidrule{2-4}
& \cellcolor{customgrey}Limited Query Processing & 
Instances where the model recognizes that further information is required for a conclusive answer, yet receives no feedback, indicating a gap in programming or query processing. & 
The model responds \textbf{'Need More Information'} for 'What is the name of the San Francisco newspaper?' but receives no further feedback. \\
\bottomrule
\end{tabular}%
}
\end{table*}

Various methodologies have been developed to enhance LLMs with KG-based RAG systems. By leveraging structured and meticulously curated knowledge from these graphs, the retrieved information is more likely to be factually accurate. 

We assessed the effectiveness of these methods and analyzed their accuracy in retrieving information for fact-based question-answering (QA) tasks using a KG. Although most of these models surpass the performance of zero-shot QA conducted directly from various standard LLMs, there is still considerable scope for improvement. For our study, we chose the WebQuestionsSP (WebQSP) \cite{yih2016value} dataset for knowledge graph question answering (KGQA), which is frequently utilized by KG-based RAG methods\cite{tan2023can}. This dataset, based on the Freebase KG \cite{bollacker2008freebase}, consists of questions that require up to two or three-hop reasoning to identify the correct answer entity, utilizing Hits@$k$ as the evaluation metric to determine if the top-$k$ predicted answer is accurate. It includes approximately 1600 test samples. The vanilla ChatGPT (GPT-3.5) accuracy in zero-shot setting without any external knowledge is 61.2\%. 

StructGPT \cite{jiang2023structgpt} is a state-of-the-art approach that leverages LLM's capabilities for reasoning with evidence extracted from a KG. This method involves extracting a sub-graph from a KG by matching the topic entities in the question. The LLM is then directly employed to identify useful relations and extract relevant triples from the sub-graph, guiding it to effectively traverse and reason within the graph structure. The Hits@1 accuracy of StructGPT on the WebQSP dataset, when utilizing ChatGPT (GPT-3.5) for question-answering tasks, was reported to be 72.6\%. In this study, we have chosen StructGPT as our reference model to analyze the current SOTA developments of KG-based RAGs in the QA setting.

We began our analysis by closely examining the failure instances of StructGPT on the WebQSP dataset. We meticulously reviewed logs from around 435 error cases to understand the model's behavior during the reasoning process. Initially, we manually analyzed 10\% of these error samples to identify common error types. Building on this manual analysis, we developed an LLM-assisted pipeline using few-shot samples to categorize the remaining errors and identify any additional error types. We then employed LLM-critic to provide recommendations and identify recurring themes based on our analysis and category mapping. These LLM-generated suggestions were subsequently validated through manual review. This detailed examination allowed us to pinpoint distinct error patterns, leading to the identification of eight primary error categories. These issues were further organized into two main divisions: \textit{Reasoning Failures}, which encompass errors arising from reasoning deficiencies, and \textit{Structural Limitations}, which include structural issues within the knowledge graph.

\noindent \textbf{Reasoning Failures:} Most failures stem from the LLMs' inability to reason correctly. These issues primarily include a failure to accurately understand the question, leading to difficulty in mapping the question to the available information. Additionally, LLMs struggle to effectively apply the cues in the question to narrow down the relevant entities. They also often fail to apply specific constraints that logically limit the search space. Generally, LLMs have difficulty grasping specifics such as temporal context, aggregating or summarizing answers, and disambiguating among multiple choices. Furthermore, they frequently choose incorrect relations, particularly in complex queries requiring multi-hop reasoning, finding it challenging to focus on the relevant elements necessary to formulate an answer. In Table~\ref{tab:err_ana}, we detail various reasoning failures, each illustrated with an example.

\noindent\textbf{Structural Limitations:} These issues occur when knowledge becomes inaccessible due to limitations in the structural design of the knowledge base or inefficient processing methods. In Table~\ref{tab:err_ana}, we categorize these challenges under structural issues, limited query processing, and the selection of inappropriate evaluation metrics.

In this work, our primary focus is on addressing errors stemming from reasoning failures in LLM models and enhancing their reasoning capabilities. Structural limitations, on the other hand, can be resolved through careful programming and the selection of more appropriate evaluation metrics. Our analysis of reasoning error samples reveals two main challenges: \textbf{(i)} Models frequently fail to grasp the question's intent, relying primarily on structural cues and semantic similarity to extract relevant relations and generate answers. \textbf{(ii)} They struggle to align the question's context with the available information.

This inability to comprehend intent and context leads to incorrect relation rankings and the misapplication of constraints. A review of response logs from both failed and successful interactions reveals that the LLM relies heavily on semantic matching. While this approach suffices for simple queries, it falls short in handling complex questions that demand multi-hop reasoning and deep contextual understanding. Therefore, improving intent identification and context alignment is essential for enhancing model performance.

\section{Mindful-RAG}
In response to our findings, we introduce \textbf{Mindful-RAG}, designed to address two critical gaps: the lack of question-intent identification and the insufficient contextual alignment with available knowledge. This approach employs a strategic hybrid method that integrates the model's intrinsic parametric knowledge with non-parametric external knowledge from a KG. The following steps provide a detailed overview of our design and methodology, each accompanied by an illustrative example.

\begin{itemize}
    \item \textbf{Step 1. Identify key Entities and relevant Tokens:} The first step is to pinpoint the key entities within a question to facilitate the extraction of pertinent information from an external KG or a sub-graph within a KG. Additionally, in our method, we task the LLM model with identifying other significant tokens that may be crucial for answering the question. For instance, consider the question from WebQSP, \textbf{``Who is Niall Ferguson's wife?"} The key entity identified by the model is \textit{`Niall Ferguson'}, and the other relevant token is \textit{`wife'}.
     \item \textbf{Step 2. Identify the Intent:} In this step, we leverage the LLM's understanding to discern the intent behind the question, prompting it to focus on keywords and phrases that clarify the depth and scope of the intent. For instance, in the provided example, the model identifies the question's intent as \textit{``identify spouse"}.
     \item \textbf{Step 3. Identify the Context:} Next, the model was instructed to understand and analyze the context of the question, which is essential for formulating an accurate response. For the provided example, the model identifies relevant contextual aspects such as \textit{``personal relationships," ``marital status," and ``current spouse."}
     \item \textbf{Step 4. Candidate Relation Extraction:} Next, the key entity relations are extracted from the sub-graph within one-hop distance. For our example, the candidate relations include information about the subject's profession, personal life, and societal role.
     \item \textbf{Step 5. Intent-Based Filtering and Contextual Ranking of Relations:} In this step, the model conducts a detailed analysis to filter and rank the extracted relations and entities based on the question's intent, ensuring relevance and accuracy. Relations are ranked according to their contextual significance, with the top-$k$ relations being selected. For example, considering the intent and context in the given scenario, the model identifies \textit{``people.person.spouse\_s"} as the most relevant relation.
    \item \textbf{Step 6. Contextual Alignment of Constraints:} In this step, the model considers temporal and geographical constraints by utilizing relevant data from various indicators to address more complex queries. This process ensures that responses are accurately tailored to specific times, locations, or historical periods. Once constraints are identified, the model aligns them contextually and refines the list of candidate entities. For example, in our scenario, the model identified constraints such as names of spouses, marriage start and end times, and the location of the ceremony. It then narrowed the list to potential spouses and extracted all related triples. Finally, the model aligned this information with the context of the `current spouse,' resulting in the correct response of \textit{`Ayaan Hirsi Ali'}, in contrast to existing methods \cite{jiang2023structgpt}, where the LLM incorrectly selected the first name on the spouse list, \textit{`Sue Douglas'}.
    \item \textbf{Step 7. Intent-Based Feedback:} In the final step, the model is prompted to validate whether the final answer aligns with the initially identified intent and context of the question. If the answer does not meet these criteria, the model is instructed to revisit \textbf{Step 5 and 6} to further refine its response.
\end{itemize}

Similarly, the model adeptly contextualizes and aggregates pertinent information in other instances. For example, when asked, \textit{``What songs did Justin Bieber write?"} it successfully compiles all relevant songs. In response to, \textit{``What is the state flower of Arizona?"} it identifies \textbf{\textit{`Arizona'}} as the key entity, with \textbf{\textit{`state'}} and \textbf{\textit{`flower'}} as relevant tokens. It correctly interprets the intent to \textbf{\textit{``identify state flower"}} and recognizes the context of \textbf{\textit{`botany,' 'state symbols,'}} and \textbf{\textit{`Arizona's official flora'}} choosing the appropriate relation: \textit{``government.governmental\_jurisdiction.official\_symbols."}
In contrast, traditional methods only identify \textit{`Arizona'} as the key entity, often missing the broader context, leading to choosing incorrect relations, \textit{``base.locations.states\_and\_provinces.country"} and answer stating the state flower of Arizona is unknown.

\textbf{Mindful-RAG} leverages the LLM's intrinsic understanding in the first three steps to identify not only the key entities but also to gather additional information such as relevant tokens, intent, and current context, all of which are essential for accurately answering the question.
These steps enable the model to appropriately filter relations and align constraints with the current context. By incorporating these steps, the LLM becomes more mindful of the specific elements to consider. In the final two steps, the LLM is prompted to tailor its response and align it with specific constraints such as time, location, and any requirements for aggregating an answer. 

\section {Experiments and Results}
\noindent\textbf{Datasets:} We evaluate our approach on two benchmark KGQA datasets, specifically WebQSP and MetaQA  \cite{zhang2018variational}. MetaQA features questions related to the movie domain, with answers up to three hops away from the topic entities in a movie KG (based on OMDb). Here, we focused only on 3-hop questions. 

\begin{figure}[t]
    \centering
    \includegraphics[width=0.48\textwidth]{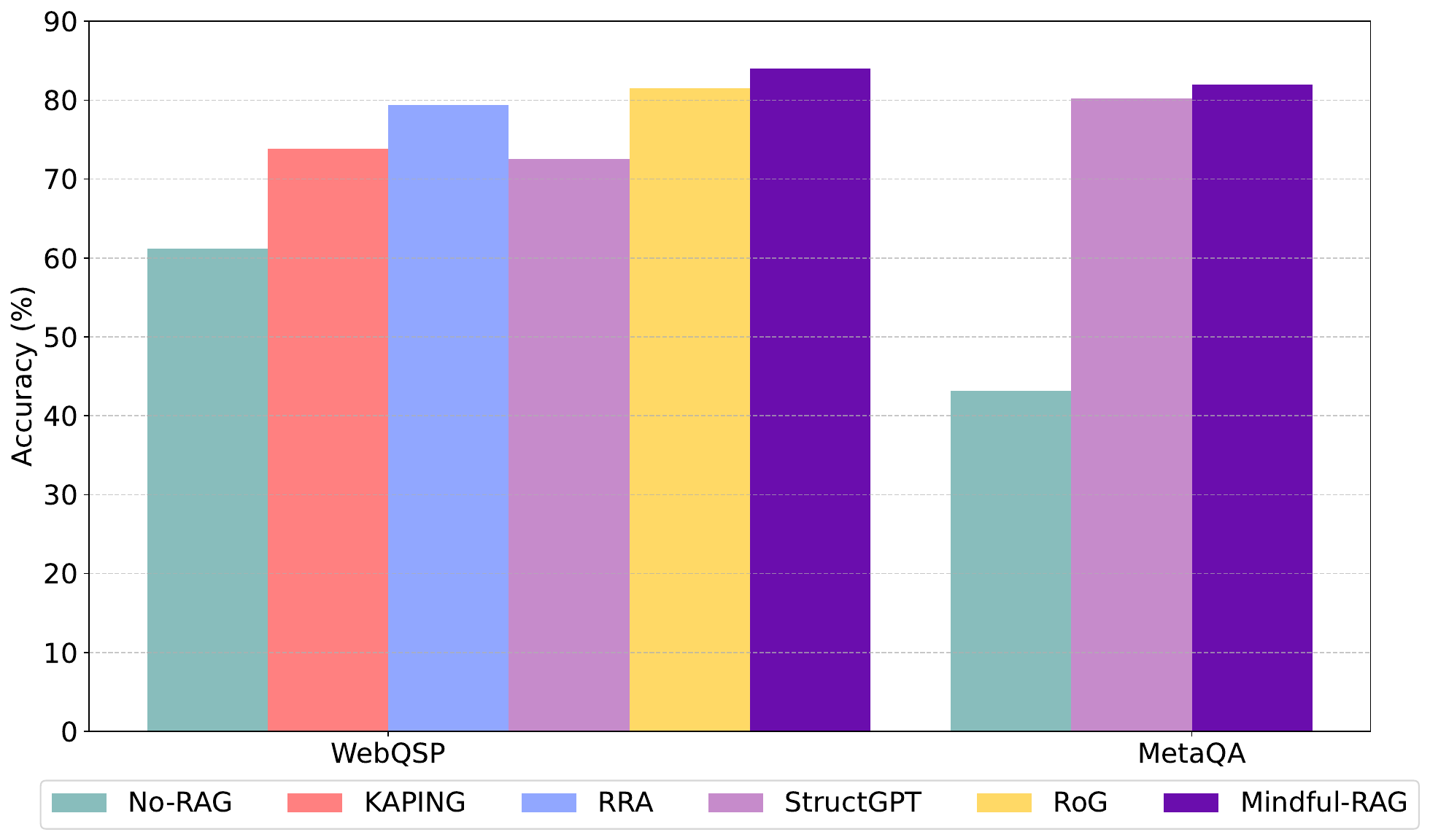}
    \caption{Mindful-RAG results on WebQSP and MetaQA}
    \label{fig:results}
\end{figure}

In our analysis of the WebQSP dataset, we evaluated several baseline methods: KAPING \cite{baek2023knowledge}, Retrieve-Rewrite-Answer (RRA) \cite{wu2023retrieve}, Reasoning on Graphs (RoG) \cite{luo2023reasoning}, and StructGPT \cite{jiang2023structgpt}. For MetaQA (3-hop), StructGPT \cite{jiang2023structgpt} served as the baseline. The results for these methods were taken directly from the respective publications. In our experiments, we adapted the base code of StructGPT \cite{jiang2023structgpt} and modified it only for improved reasoning as outlined in the previous section. We also examined the performance of ChatGPT without RAG on both datasets. The results, presented in Figure \ref{fig:results}, show that Mindful-RAG, shows promising improvement in accuracy of reasoning error cases achieving a Hits@1 of 84\% on WebQSP and 82\% on MetaQA. Additional accuracy improvements can be achieved by addressing structural issues and incorporating partial answers to enhance precision, rather than relying solely on exact matches.

The primary goal of this study is to explore methods for mitigating reasoning errors in KG-RAG systems.  It is important to emphasize that our approach serves as an initial demonstration of the potential in combining the parametric knowledge of models with non-parametric external knowledge. More advanced RAG methods could be developed in the future to significantly surpass the performance of our approach.


\section{Related Work}
Recent efforts to enhance RAG systems have focused on various improvements. Siriwardhana et al. \cite{siriwardhana2023improving} aimed to improve domain adaptation for Open Domain Question Answering (ODQA) by jointly training the retriever and generator and enriching the Wikipedia-based knowledge base with healthcare and news content. RAFT \cite{zhang2024raft} enhances RAG by customizing language models for specific domains in open-book QA. Self-RAG \cite{asai2023self} aims to increase the factual accuracy of LLMs through adaptive self-critique and retrieval-generation feedback loops. Fit-RAG \cite{mao2024fit} introduces a method that uses detailed prompts to ensure deep question understanding and clear reasoning in fact retrieval. Domain-specific knowledge graphs \cite{abu2021domain, agrawal2022building, tang2023construction, agrawal2023aiseckg} have been effectively employed in KG-based RAG systems within LLMs \cite{delile2024graph, jianghykge, agrawal2024cyberq} for question-answering tasks \cite{zhao2024lb,agrawal2023auction}. While most efforts focus on enhancing LLMs by augmenting knowledge graphs with relevant facts, there has been limited work on improving the reasoning capabilities of LLMs during knowledge retrieval. Our research with Mindful-RAG aims to establish a road map for advancing these methods by leveraging the model’s inherent knowledge for better question understanding.

\section{Discussion and Conclusion}
We conducted an error analysis of KG-based RAG methods integrated with LLMs for question-answering tasks, identifying eight critical failure points, categorized into reasoning failures and structural limitations. Reasoning failures involve LLMs struggling with understanding questions and leveraging contextual clues, particularly in cases involving temporal context and complex relational reasoning. Structural limitations pertain to inadequate attention to the structure of the knowledge base and weaknesses in evaluation metrics. These challenges highlight areas for improvement, especially in handling complex, multi-hop queries. To address these issues, we propose Mindful-RAG, designed to enhance intent-driven retrieval and ensure contextually coherent responses, directly targeting the identified deficiencies. While our approach focuses on mitigating reasoning-based failures, future research could explore addressing structural issues through the use of feedback and human-in-the-loop input. Additionally, combining vector-based search with KG-based sub-graph retrieval represents a promising direction for enhancing LLM performance in knowledge-intensive tasks.

\section*{Acknowledgments}
This material is based upon work supported by
the National Science Foundation under Grant No.
2335666.

\bibliography{reference}
\bibliographystyle{IEEEtran}

\end{document}